\documentclass[aps,prb,twocolumn,showpacs,floatfix,superscriptaddress,tightenlines,amsmath,amssymb]{revtex4-2}
\pdfoutput=1

\usepackage[T1]{fontenc}
\usepackage{natbib}
\usepackage{graphicx}
\graphicspath{{figures/}}
\usepackage{amscd}
\usepackage{amsmath,amsfonts,amssymb,mathrsfs}
\usepackage{gensymb}
\usepackage{dcolumn}
\usepackage{bm}
\usepackage{float}
\usepackage[bookmarks=false]{hyperref}
\hypersetup{hidelinks,breaklinks,colorlinks,citecolor=blue,linkcolor=black,urlcolor=blue}
\usepackage{textcomp}
\usepackage{array}
\usepackage{booktabs}
\usepackage{xspace}

\newcommand{\nro}{{NiRh$_2$O$_4$}\xspace}

\let\oldcdot=\cdot
\def\cdot{\negthinspace\oldcdot\negthinspace}

\begin{document}

\title{\texorpdfstring{Electronic structure of the frustrated diamond lattice magnet \nro}{}}

\author{B. Zager}
\affiliation{Department of Physics, Brown University, Providence, Rhode Island 02912, United States}
\author{J. R. Chamorro}
\affiliation{Department of Chemistry, Johns Hopkins University, Baltimore, MD, USA}
\affiliation{Institute for Quantum Matter, William H. Miller III Department of Physics and Astronomy, Johns Hopkins University, Baltimore, MD, USA}
\author{L. Ge}
\affiliation{School of Physics, Georgia Institute of Technology, Atlanta, Georgia 30332, USA}
\author{V. Bisogni}
\affiliation{National Synchrotron Light Source II, Brookhaven National Laboratory, Upton, New York 11973, USA}
\author{J. Pelliciari}
\affiliation{National Synchrotron Light Source II, Brookhaven National Laboratory, Upton, New York 11973, USA}
\author{J. Li}
\affiliation{National Synchrotron Light Source II, Brookhaven National Laboratory, Upton, New York 11973, USA}
\author{G. Fabbris}
\affiliation{Advanced Photon Source, Argonne National Laboratory, Lemont, Illinois 60439, USA}
\author{T. M. McQueen}
\affiliation{Department of Chemistry, Johns Hopkins University, Baltimore, MD, USA}
\affiliation{Institute for Quantum Matter, William H. Miller III Department of Physics and Astronomy, Johns Hopkins University, Baltimore, MD, USA}
\affiliation{Department of Materials Science and Engineering, Johns Hopkins University, Baltimore, MD, USA}
\author{M. Mourigal}
\affiliation{School of Physics, Georgia Institute of Technology, Atlanta, Georgia 30332, USA}
\author{K. W. Plumb}
\affiliation{Department of Physics, Brown University, Providence, Rhode Island 02912, United States}

\date{\today}

\footnotetext[1]{See Supplemental Material} 

\begin{abstract}

The $A$-site spinel \nro{} is the only known realization of a spin-1 diamond lattice magnet and is predicted to host unconventional magnetic phenomena driven by frustrated nearest and next-nearest neighbor exchange as well as orbital degeneracy. Previous works found no sign of magnetic order but found a gapped dispersive magnetic excitation indicating a possible valence bond magnetic ground state. However, the presence of many competing low energy degrees of freedom and limited empirical microscopic constraints complicates further analysis. Here, we carry out resonant inelastic x-ray scattering (RIXS) and x-ray absorption spectroscopy (XAS) to characterize the local electronic structure of \nro{}. The RIXS data can be partly described by a single-ion model for tetrahedrally coordinated Ni$^{2+}$ and indicates a tetragonal distortion $\Delta t_2\!=\!70$~meV that splits the $t_2$ orbitals into a high energy orbital singlet and lower energy orbital doublet. We identify features of the RIXS spectra that are consistent with a Rh-Ni two-site excitation indicating strong metal-metal hybridization mediated by oxygen in \nro{}. We also identify signatures of electron-phonon coupling through the appearance of phonon sidebands that dress crystal field excitations. These results establish the key energy scales relevant to the magnetism in \nro{} and further demonstrate that covalency and lattice dynamics play essential roles in controlling the magnetic ground states of $A$-site spinels. 

\end{abstract}

\maketitle

\section{Introduction} \label{intro}

Materials with frustrated magnetic interactions provide a platform for realizing novel phases which avoid conventional symmetry-breaking order \cite{lacroix2011introduction}. Such phases are acutely sensitive to a hierarchy of competing energy scales involving spin, charge, orbital, and lattice degrees of freedom \cite{khomskii2014transition}. This enables precise tuning of the collective orders in these systems to explore new phenomena and develop new technologies \cite{basov2017towards}. Along this line, there has recently been a significant interest in novel phases driven by strong spin orbit coupling (SOC) that has motivated significant work on $4d$/$5d$ transition metal compounds \cite{takayama2021spinorbit}. In some cases, however, the relatively weak SOC in $3d$ compounds may still become relevant. In particular, when there is an orbital degeneracy, SOC may compete with the Jahn-Teller (JT) effect to produce a spin-orbital-lattice entangled state \cite{streltsov2020JTandSOC}. 

The spinel structure comprises an intensely studied class of materials with both fundamental significance and widespread applications. The general formula $AB_2X_4$ contains tetrahedrally coordinated $A^{2+}$ ions and octahedrally coordinated $B^{3+}$ ions which occupy diamond and pyrochlore sublattices respectively. In the case of magnetic $B$ ions, the system forms a prototypical geometrically frustrated pyrochlore magnet. On the other hand, for magnetic $A$ site ions, the bipartite diamond lattice may be frustrated in the presence of competing nearest and next-nearest neighbor antiferromagnetic exchange \cite{tsurkan2021spinels}. Previously studied diamond lattice magnets include the $A$-site spinels Co$_3$O$_4$ \cite{roth1964co3o4}, $A$Al$_2$O$_4$ ($A=\;$Mn, Fe, Co) \cite{tristan2005geometric,krimmel2009spin}, $A$Sc$_2$S$_4$ ($A=\;$Mn, Fe) \cite{fritsch2004spin,gao2017spiral}, MnSc$_2$Se$_4$ \cite{guratinder2022magnetic}, and $A$Rh$_2$O$_4$ ($A=\;$Co, Cu, Ni) \cite{ge2017spin,chamorro2018frustrated}, as well as the lanthanides LiYbO$_2$ \cite{bordelon2021frustrated} and NaCeO$_2$ \cite{bordelon2021afm}. These materials host a variety of magnetic phenomena, ranging from long-range ordered states to disordered spin liquid and spin glass states. In many cases, especially in the presence of an orbital degree of freedom, these materials lie near a quantum critical point, with multiple competing phases exhibiting strong sensitivity to disorder and external perturbations \cite{chen2009spinorbital,savary2011impurity,nair2014approaching,macdougall2016revisiting,plumb2016afm,biffin2017magnetic,tsurkan2017structure,naka2020chemical,cho2021pressure}. It is thus essential to characterize the microscopic energy scales associated with the spin, lattice, and orbital degrees of freedom in these materials.


Another topic of recent interest is frustrated magnetism in spin-1 materials \cite{li2018competing,liu2020featureless,morey2019ni2mo3o8}, which support more degrees of freedom than spin-$\frac{1}{2}$, while still being sensitive to quantum fluctuations. In particular, recent studies have predicted spin-1 diamond lattice antiferromagnets to host unconventional magnetic phenomena, namely topological paramagnetism \cite{wang2015topological}, spiral spin liquid phases \cite{buessen2018quantum}, quantum critical phenomena \cite{chen2017quantum}, and excitonic magnetism \cite{li2019spinorbital}. 

The spinel \nro{} contains spin-1 Ni$^{2+}$ ions on the $A$ sites and nonmagnetic Rh$^{3+}$ ions on the $B$ sites, thus realizing the only known spin-1 on a diamond lattice. Early studies identified a cubic to tetragonal structural transition \cite{horiuti1964tetragonal} and an apparent antiferromagnetic transition \cite{blasse1963afm}. However, a recent study on high quality samples found no sign of magnetic order down to 0.1~K \cite{chamorro2018frustrated}, implying that chemical disorder may have stabilized ordering found in the original studies. In that study, specific heat and x-ray diffraction measurements found a small entropy loss associated with the structural distortion at $T=440$~K, indicating only a partial lifting of orbital degeneracy. This finding is consistent with the $p_{\text{eff}} = 3.3\mu_B$ paramagnetic moment that is significantly larger than the pure spin-1 value of $2.83\mu_B$, implying an orbital contribution to the magnetism. The Curie-Weiss temperature of $\Theta_{\text{CW}} = -11$~K indicates a large frustration parameter $f = \Theta_{\text{CW}}/T_N > 100$. Inelastic neutron scattering (INS) found gapped dispersive magnetic excitations suggestive of a valence bond ground state. However, an incomplete knowledge of the electronic ground state has limited modeling of the INS data. Recent theoretical work predicted that spin-orbit coupling, crystal fields, and correlations generate a spin-orbital singlet ground state in \nro{} \cite{li2019spinorbital,das2019nirh2o4}. Such a state would explain the absence of magnetic ordering and key features of the magnetization, specific heat, and neutron data. But measurements of the high energy excitations characterizing the orbital configurations of \nro{} are required to confirm this picture.  

In this paper, we use a combination of resonant inelastic x-ray (RIXS), x-ray absorption spectroscopy (XAS), and inelastic neutron scattering to provide a detailed account of the electronic structure of \nro{}. We observe crystal field excitations for  Ni$^{2+}$ in a distorted tetrahedral environment that are well-described by a single-ion model including Coulomb interaction, spin-orbit coupling, crystal field splitting. We find additional electronic excitations corresponding to two-site Rh to Ni charge transfer. We also find that the orbital excitations are coupled to optical phonons giving rise to distinct phonon sidebands that dress $d$-$d$ excitations. These results provide a detailed description of the local electronic structure of a novel frustrated magnet along with the key energy scales required for understanding the magnetic ground state and low-energy excitations. We have additionally shown that both covalency and lattice dynamics play essential roles in this material and should be considered in any realistic model for the magnetism. These insights provide guidance for exploring novel effects in \nro{} and other magnetically frustrated spinels using pressure, magetic fields, and chemical substitution. 

This paper is organized as follows. In section \ref{experiment}, we describe the experimental details. In section \ref{results}, we present the RIXS, XAS, and INS results and describe our attempts to model the data. We first use a single ion model for Ni$^{2+}$ in a distorted tetrahedral crystal field, then a minimal two-site Ni-Rh hopping model, and finally we incorporate electron phonon coupling in order to capture the lineshapes of the RIXS spectra. Our findings are summarized in section \ref{conclusion}. 
 
\begin{figure*}[t!]
   \begin{center}
   \includegraphics{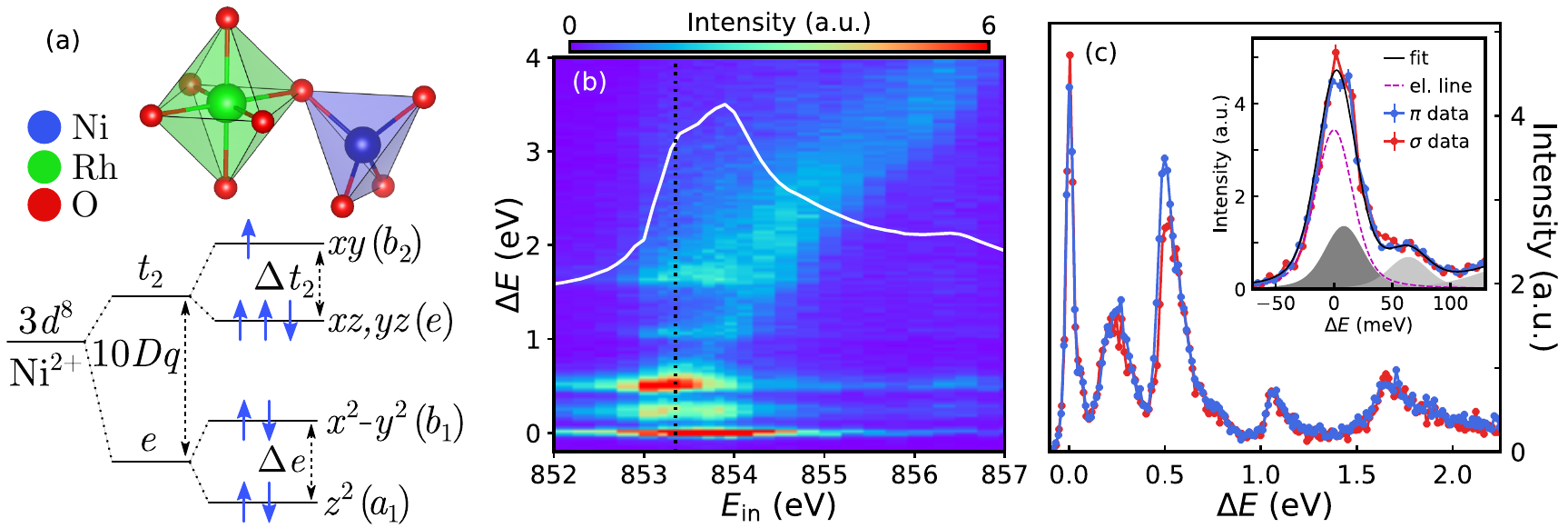}
   \caption{Overview of RIXS data measured at $T\!=\!40$~K and $2\theta\!=\!90$\degree{}.
   (a) Local coordination of Ni and Rh in \nro{}. Rh is octahedrally coordinated while Ni$^{2+}$ $3d$ orbitals are split by tetrahedral crystal field with tetragonal distortion $\Delta t_2\!>\!0$. (b) Measured RIXS intensity vs. energy loss and incident energy measured in $\pi$ polarization. The XAS measured in total fluorescence yield is plotted in white. (c) RIXS spectrum in both $\pi$ and $\sigma$ polarization at $E_{in}\!=\!853.4$~eV, indicated by the dashed line in (b). The inset shows the low-energy part of the spectrum and an empirical fit.}
   \label{fig:rixs}
   \end{center}
\end{figure*}

\section{Experiment} \label{experiment}

Polycrystalline \nro{} was synthesized following the methods described in \cite{chamorro2018frustrated}. The samples contain 4\% nonmagnetic Rh$_2$O$_3$ by mass, which is not expected to contribute any observable effects to our measurements. 

Resonant inelastic x-ray scattering (RIXS) measurements at the Ni $L_3$ edge were performed at the soft inelastic x-ray scattering (SIX) beamline 2-ID at the National Synchrotron Light Source II (NSLS-II) at Brookhaven National Laboratory. Measurements were carried out  at 40~K and 25~K using an incident x-ray polarization in both linear horizontal ($\pi$) and linear vertical ($\sigma$) geometries. The scattering angle was fixed at 90{\degree} to minimize the contribution from Thomson elastic scattering and giving a momentum transfer of $Q \! =\! 0.61$~\AA$^{-1}$. The incident x-ray energy was varied across the Ni~$L_3$ edge ($\sim$853 eV). The combined energy resolution was determined to be 31~meV based on the full width at half maximum (FWHM) of the elastic signal from carbon tape.

X-ray absorption spectroscopy (XAS) measurements at the O~$K$ edge were performed at NSLS-II beamline 2-ID (SIX), measured in fluorescence yield (FY). XAS measurements at the Rh~$L$ edges at Advanced Photon Source 4-ID-D, measured in total electron yield (TEY). 

Inelastic neutron scattering (INS) measurements were performed on the Fine Resolution Fermi chopper spectrometer (SEQUOIA) at the Spallation Neutron Source, Oak Ridge National lab. The incident neutron energy was fixed at 160~meV using the coarse resolution chopper (FC2) rotating at 600~Hz. The same 4~g sample used in \cite{chamorro2018frustrated} was loaded in an aluminum can, and held at 3.6~K for the measurement. Scattering contributions from the sample environment were removed by subtraction of an empty can data set during data reduction.

\section{Results} \label{results}
Fig.~\ref{fig:rixs}(b) shows the RIXS intensity as a function of incident energy $E_{\text{in}}$ and energy loss $\Delta E$ measured at 40~K with incident $\pi$ polarization. The XAS, shown in white, contains a weak pre-edge feature at 852.9~eV, a main peak split into two features at 853.4~eV and 853.8~eV, and a satellite peak at 856.4~eV. The RIXS spectrum shows an elastic line at $\Delta E \!= \! 0$~eV, five Raman-like features at 0.065~eV, 0.25~eV, 0.5~eV, 1.1~eV, and 1.6~eV, a broad charge-transfer (CT) background between 2 and 4~eV, and a fluorescence line (FL) at constant scattered photon energy. 

Fig.~\ref{fig:rixs}(b) shows RIXS scans at the main resonance of $E_{\text{in}} \!= \! 853.4$~eV, indicated by the dotted line in Fig.~\ref{fig:rixs}(a), in both $\pi$ and $\sigma$ incident polarization at 40~K. The inset shows the low energy region near the elastic line. The elastic intensity due to Thomson scattering is expected to be strongly suppressed in the $\pi$-polarized data for the 90$\degree${} scattering angle configuration of these measurements. However, the $\pi$ and $\sigma$-polarized data show similar intensity around $\Delta E\! = \! 0$, suggesting unresolved low energy excitations. The elastic line cannot be fit by a single resolution-limited Voigt function, further indicating the presence of unresolved low energy contributions. This can be explained by the $\sim$12~ meV dispersive magnetic excitation observed in inelastic neutron scattering \cite{chamorro2018frustrated}. By including an additional resolution-limited Voigt function at 12~meV, we obtain an adequate fit to the quasi-elastic line, as shown in the inset of Fig.~\ref{fig:rixs}(b) for the $\pi$-polarized data, which also includes contributions from two overlapping higher energy peaks. A similar fit can be obtained for the $\sigma$-polarized data, with a slightly larger contribution from the elastic peak likely originating from Thomson scattering that contributes in that geometry.

\subsection{Single-ion model} \label{single}
\begin{figure}[!ht]
   \begin{center}
   \includegraphics{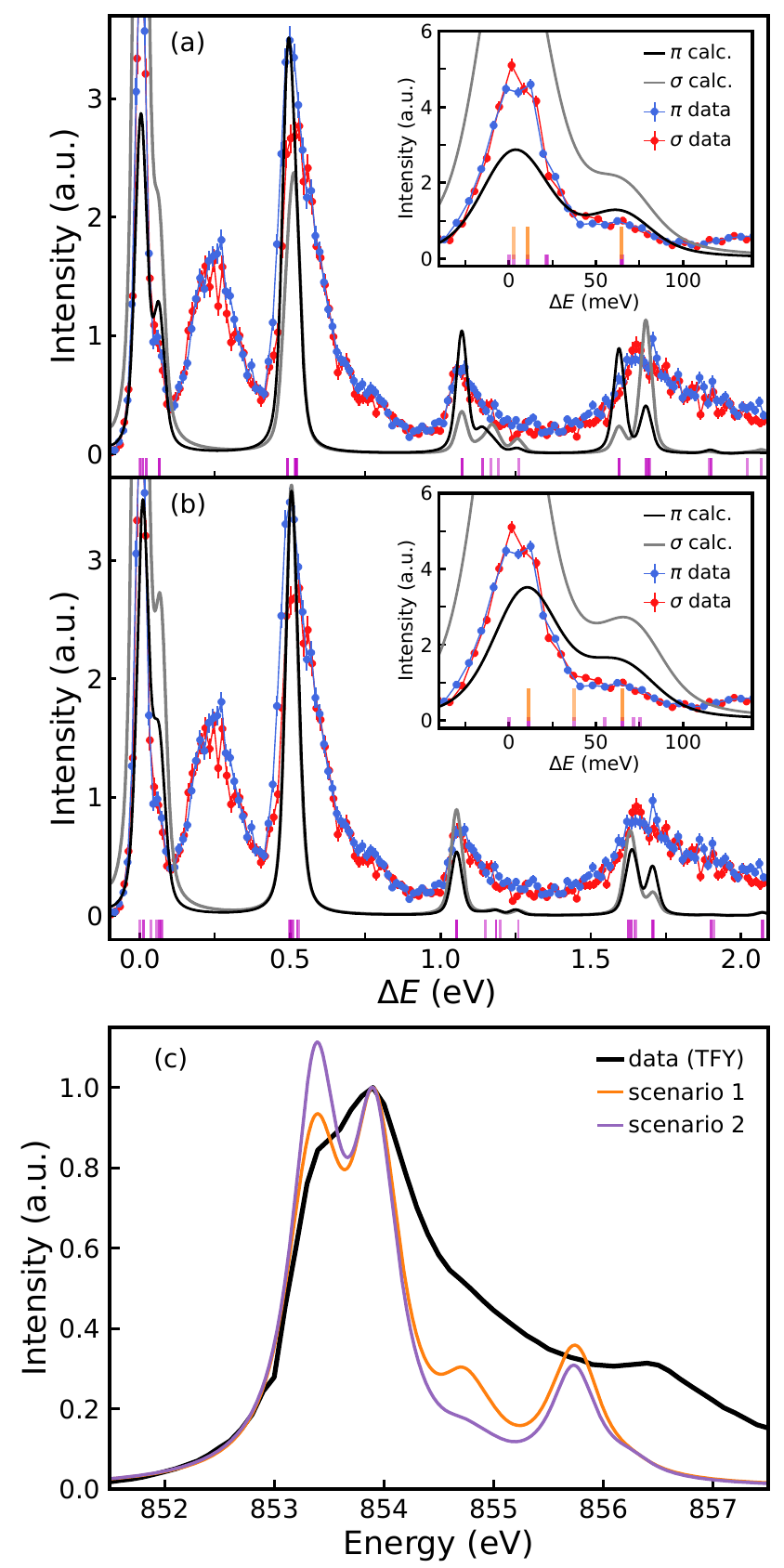}
   \caption{Comparison of the two crystal field schemes used for modeling the RIXS data. Scenario 1 ($\Delta t_2 \!> \! 0$) is shown in (a) and scenario 2 ($\Delta t_2 \!<\! 0$) is shown in (b).  The pink markers show the eigenvalues. The orange markers in the inset indicate which eigenvalues are dipole-allowed transitions from the ground state. (c) Comparison of the two schemes for modeling the Ni $L_3$ XAS (TFY). The calculated XAS spectra are normalized to the intensity at the maximum of the measured spectrum. 
   }
   \label{fig:rixs_model}
   \end{center}
\end{figure}

To identify the features in the measured spectra, we calculate the RIXS cross section for a single Ni$^{2+}$ ion including Coulomb interaction, spin-orbit coupling (SOC), and crystal field (CF) splitting. Calculations were implemented using the EDRIXS package \cite{wang2019edrixs}, further details are described in the Supplemental Material \cite{Note1}. We include onsite Coulomb interactions as parameterized by the Slater integrals: $F_{dd}^2$, and $F_{dd}^4$ describe the direct Coulomb repulsion between $d$ electrons, $F_{pd}^2$ describes direct Coulomb repulsion between $d$ electrons and the $2p$ core hole, and $G_{pd}^1$ and $G_{pd}^3$ describe the Coulomb exchange between $d$ electrons and the $2p$ core hole. Initial values for the above parameters are calculated for a free Ni$^{2+}$ ion in $3d^8$ and $2p^53d^9$ configurations by the Hartree-Fock method using Cowan's code \cite{cowan1981theory}. To account for the reduced intra-atomic repulsion due to covalency effects in the solid, we include empirical scale factors on $F_{dd}$, $F_{pd}$, and $G_{pd}$ that are determined by fitting to the RIXS data. 

The crystal-field is parameterized by the tetrahedral splitting $10Dq\!<\!0$ and the splitting due to the tetragonal distortion $\Delta t_2\!=\!E_{b_2}\!-\!E_{e'}$, $\Delta e\!=\!E_{b_1}\!-\!E_{a_1}$. The tetrahedral crystal field ($T_d$ point group) splits the $d^8$ configuration into $e^4t_2^4$. The tetragonal distortion ($T_d \rightarrow D_{2d}$) further splits the $e$ doublet into $a_1$ ($d_{z^2}$) and $b_1$ ($d_{x^2-y^2}$) singlets, and the $t_2$ triplet into a $b_2$ ($d_{xy}$) singlet and $e'$ ($d_{xy}$, $d_{yz}$) doublet, as shown in Fig.~\ref{fig:rixs}(a). These symmetry considerations leave the sign and magnitude of the splittings unconstrained. There have been no reported direct measurements of the crystal field splitting in \nro{}, and there are conflicting reports of the sign of the tetragonal splitting. This splitting cannot be constrained only from knowledge of the crystal structure because expectations from electrostatic considerations are often inaccurate due to additional effects from covalency and spin-orbit coupling \cite{khomskii2014transition}. As we will show below, both effects are significant in \nro{}.

The low temperature structure reported in \cite{chamorro2018frustrated} is tetragonal with elongated NiO$_4$ tetrahedra and compressed RhO$_6$ octahedra. This led the authors to propose a crystal field scheme with the $t_2$ levels split into a lower $b_2$ state and upper $e'$ states ($\Delta t_2\! <\! 0$), following the model from \cite{gutlich1984mossbauer}, as expected from electrostatic considerations. However, the DFT calculation in \cite{das2019nirh2o4} suggests that the $t_2$ levels are split into lower $e'$ states and an upper $b_2$ state ($\Delta t_2\! >\! 0$). We find that although a single ion model is not sufficient to explain the full RIXS spectra for \nro{}, it can capture many of the essential features, and enables us to distinguish between the two possible scenarios for the tetragonal splitting and place constraints on its magnitude.

We constrain our model to maintain consistency with material trends for insulating Ni compounds and with other spectroscopic measurements on \nro{} \cite{gutlich1984mossbauer,kocsis2013magnetoelasticity}. The peaks around 0.5~eV and 1~eV fix the value of $10Dq \approx 0.55$~eV, as they correspond to $e_{}^4t_2^4 \rightarrow e_{}^3t_2^5$ and $e_{}^4t_2^4 \rightarrow e_{}^2t_2^6$ transitions. This $10Dq$ value is consistent with DFT \cite{das2019nirh2o4} and the values for tetrahedrally coordinated Ni$^{2+}$ in other compounds \cite{gutlich1984mossbauer}. We can also constrain the value of atomic spin-orbit coupling $\lambda$ by the requirement of a dipole-allowed level near 11~meV to agree with the neutron scattering data. Finally, the peak at 1600~meV corresponds to an excitation within the Hund's multiplet and fixes the value of $F_{dd}$. To match the energy of the highest energy peak, $F_{dd}$ must be set to 0.5, giving a Hund's coupling of $J_H = \frac{1}{14}(F_{dd}^2 + F_{dd}^4) = 0.71$~eV. This parameter also sets the energy of intra-$t_2$ spin-flip excitations ($S=1 \rightarrow S=0$) between 1.1 and 1.3~eV. These excitations have minimal intensity in the computed RIXS cross section and we do not expect to observe them above the signal-to-noise of our data, see Fig.~\ref{fig:rixs_model}(a) and \ref{fig:rixs_model}(b). The 50\% reduction in the intra-atomic Coulomb interaction $F_{dd}$, compared with the atomic values, suggests strong Ni-O covalency \cite{degroot1994xas} and already indicates the inadequacy of a single ion model for \nro{}, we address this later in section \ref{hybridization}. The Slater integrals $F_{pd}$ and $G_{pd}$ determine the intermediate state energies and do not affect the RIXS peak energies. However, they do influence the RIXS intensities and energy splitting of the main XAS peak. 
\begin{table}[t!]
	\centering
	\caption{Parameters for the single-ion model using $\Delta t_2>0$ and $\Delta t_2<0$. $10Dq$, $\Delta t_2$, $\Delta e$, and $\lambda$ are in meV. $F_{dd}$, $F_{pd}$, and $G_{pd}$ are dimensionless.}
	\begin{tabular}{c c c c c c c c c}
	\hline\hline
	{}             & $10Dq$ & $\Delta t_2$ & $\Delta e$ & $\lambda$ & $F_{dd}$ & $F_{pd}$ & $G_{pd}$ \\ \hline
	$\Delta t_2>0$ & -580   & 70           & 56         & 13        & 0.5      & 0.7     & 0.75 \\
	$\Delta t_2<0$ & -530   & -50          & -40        & 27        & 0.5      & 0.7     & 0.75 \\
	\hline\hline
	\end{tabular}
	\label{tab:parameters}
\end{table}
We have tested models for tetrahedral compression and elongation and find that they both capture many features of the data; table \ref{tab:parameters} shows the best parameters for the two scenarios (1) $\Delta t_2\!>\! 0$ and (2)  $\Delta t_2\!<\! 0$, plotted in Fig.~\ref{fig:rixs_model}. $\Delta e$ is underdetermined by our data and mainly contributes to the splitting of the peaks above 500 meV. Here, we assume that $\Delta e$ has the same sign as $\Delta t_2$. We also expect $|\Delta e| < |\Delta t_2|$, due to the reduced Ni-O hybridization of the $e$ states  compared to the $t_2$ states in tetrahedral symmetry. We find that the constraint $\Delta e = 0.8\Delta t_2$ provides an adequate agreement with the data as shown in Fig.~\ref{fig:rixs_model}. Both models reproduce the observed peaks at 65~meV, 0.5~eV, and 1.5~eV and while scenario 1 better reproduces the small dichroism of the 0.5~eV feature, scenario 2 more faithfully reproduces the negligible dichroism of the 1~eV and 1.6~eV peaks. The high energy excitations, $\Delta E\!>\!500$~meV, are far above the insulating gap in \nro{} and are thus coupled to delocalized states. This effect is not captured in the single-ion model and may explain the Fano-like lineshape. However, we found that a more careful examination of the low energy excitations for $E<100$~meV enables a distinction between crystal field models. 

The low energy part of the spectrum is shown in the insets of Fig.~\ref{fig:rixs_model}(a) and \ref{fig:rixs_model}(b), with the orange markers indicating energy levels with nonvanishing neutron cross section. Since photons emitted with energies close to the absorption edge are more likely to be reabsorbed, we expect strong self absorption effects near the elastic line and our model should predict a larger quasi-elastic intensity than what is observed. The observed intensities are also likely modified by intersite magnetic exchange interactions \cite{das2019nirh2o4} that are not included in our single site model. Nevertheless, a careful comparison of the observed linear dichroism at low energy transfers reveals that scenario 1, $\Delta t_2 \!>\!0$, is more consistent with our data. The low energy subspace of scenario 1 is equivalent to the model from \cite{das2019nirh2o4}, and yields excited states at 3$^*$, 11(2)$^*$, 22(2), and 65(3)$^*$~meV. The low energy subspace of scenario 2 is equivalent to the model from \cite{gutlich1984mossbauer}, yielding excited states at 11(2)$^*$, 38$^*$, 55, 65(2)$^*$, 72, and 75~meV. Asterisks indicate those states with nonvanishing neutron intensity and parentheses indicate the state degeneracy, disregarding any splitting of less than 1~meV. The calculation for scenario 1 shows that the dominant contribution to the quasi-elastic RIXS intensity comes from the ground state and a 3~meV excitation for both polarizations [Fig.~\ref{fig:rixs_model}(a)]. This is consistent with the nearly equivalent quasielastic lines we measured for each polarization. However, for scenario 2, the dominant contribution comes from the ground state for $\sigma$ polarization and from the 11~meV excitation for $\pi$ polarization [Fig.~\ref{fig:rixs_model}(b)]. Although the energies of these excitations fall within our experimental resolution, scenario 2 should result in a more pronounced difference in the quasielastic lineshape between $\sigma$ and $\pi$ polarizations and that is not consistent with our data.

To provide an additional check for our single ion model, we compare with the measured x-ray absorption spectrum (XAS) at the Ni $L_3$ edge measured in total fluorescence yield (TFY) shown in Fig.~\ref{fig:rixs_model}(c). The main XAS peak is split into lower and upper peaks at 853.4 and 853.8 eV, corresponding to the states $2p^5 e^4t_2^5$ and $2p^5 e^3t_2^6$. In the single-ion model, this splitting depends on $10Dq$, $F_{pd}$, and $G_{pd}$. Although the XAS lineshape is known to be distorted by TFY measurements, the calculated spectrum shows better qualitative agreement with our data for scenario 1. In particular, a model with $\Delta t_2 > 0$ more faithfully reproduces the relative intensities of the 853.4 and 853.8 eV peaks, and captures additional observed intensity at 854.7~eV that is not predicted by a model with $\Delta t_2 < 0$ [Fig~\ref{fig:rixs_model} (c)]. However, we find that for either scenario, the single-ion model cannot capture the large, $\sim 3$~eV, splitting between the main peak and satellite peak at 856.4~eV. 

Based on the above considerations, we conclude that scenario 1, splitting the $t_2$ levels into lower energy $d_{xz},d_{yz}$ orbitals and a higher energy $d_{xy}$ orbital, is the correct crystal field scheme in \nro{}. However, there are many notable discrepancies between the single ion model and our data. First, a single-ion model completely fails to reproduce the intense 250~meV RIXS peaks for any reasonable set of parameters \cite{Note1}, and second, it does not accurately capture the broad asymmetric lineshape of the 0.5~eV peak. Both of these features in the RIXS spectra arise because of two distinct effects that cannot be accounted for in a single-ion description. First, there is strong Ni-Rh hybridization, and second, there is strong electron-phonon coupling. The evidence for each of these effects and a detailed discussion of their respective influence on the RIXS spectra and magnetism in \nro{} is discussed below in sections \ref{hybridization} and \ref{epc}.

\subsection{Ni\hspace{1pt}-Rh hybridization} \label{hybridization}
\begin{figure}[!t]
   \begin{center}
   \includegraphics{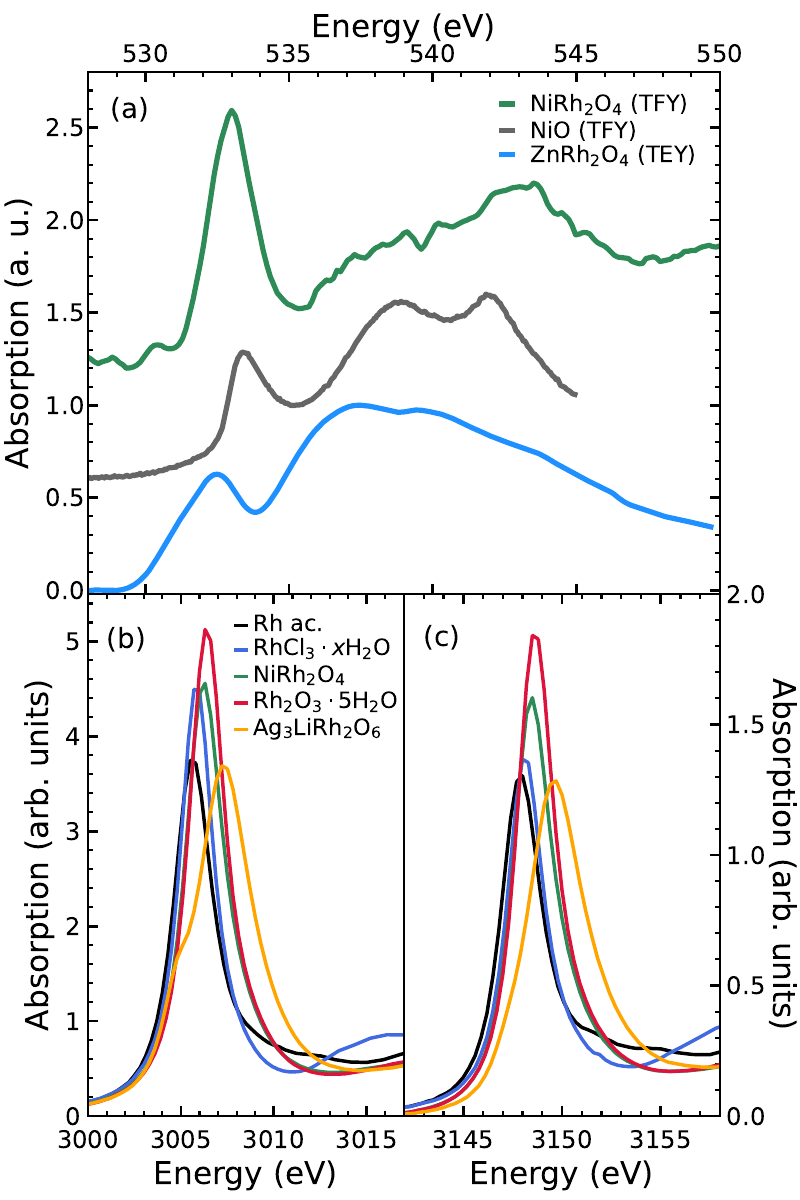}
   \caption{(a) XAS at the O $K$ edge for \nro (TFY), with NiO (TFY) and ZnRh$_2$O$_4$ (TEY) for comparison (NiO data from \cite{kurmaev2008oxygen} and ZnRh$_2$O$_4$ data from \cite{wahila2019zro}) (b) XAS measured at the Rh $L_3$ edge and (c) the Rh $L_2$ edge.}
   \label{fig:xas}
   \end{center}
\end{figure}

We now extend our model to include the influence of electronic hybridization between Ni and Rh sites. Such metal-metal hybridization is supported both by experiment and ab-initio calculations. Density functional theory calculations \cite{das2019nirh2o4} have found strong Ni-Rh hybridization is mediated by the intermediate oxygens. Furthermore, the calculation found an insulating gap of 250~meV, with the highest occupied and lowest unoccupied states having mostly Rh and Ni character respectively. This suggests that the lowest interband transition consists of Rh$\rightarrow$Ni excitations.

We have carried out x-ray absorption (XAS) measurements at the O $K$ and Rh $L$ edges in order to characterize the electronic states at the Rh and O sites in \nro{}. Fig.~\ref{fig:xas}(a) shows the O~$K$ edge XAS for \nro{} measured in total fluorescence yield (TFY), with the spectra for NiO (TFY) and ZnRh$_2$O$_4$ (TEY) shown for comparison. NiO provides a comparison to bonding in an NiO$_6$ octahedron while ZnRh$_2$O$_4$, being isostructural to the cubic phase of \nro, provides a comparison to bonding in a RhO$_6$ octahedron in the absence of an unfilled neighboring $3d$ shell.


The pre-edge region contains a small peak at 530.5 eV and a large peak at 533 eV, corresponding to O $2p$ states hybridized with empty metal $d$ states. The region above 535~eV corresponds to Ni $4sp$ and Rh $5sp$ states. By comparing to the projected density of states from DFT calculations \cite{das2019nirh2o4}, we can assign the small pre-edge peak at 530.5~eV to the unfilled Ni $t_2$ states and the large peak at 533~eV to the unfilled Rh $e_g$ states. The pre-edge peak intensity is determined by both the number of empty metal states and the degree of hybridization \cite{frati2020oxygen}. In NiO and ZnRh$_2$O$_4$, where the metal sites provide two empty states per O site, the pre-edge peak intensities are comparable. In \nro{}, there are 2 empty Rh states and 0.5 empty Ni states per O site. This small increase in the number of available states alone cannot explain the significant enhancement of pre-edge peak intensity. The intensity can thus be explained by an increased hybridization due to the cooperative influence of the Ni and Rh. This can be expected based on the large inductive effect of Rh$^{3+}$ \cite{lenglet2000ligand}. It may also be enhanced due to the long range exchange interactions. In the five-site exchange pathway $A$-O-$B$-O-$A$ via a nonmagnetic $B$ cation, the dominant contribution is thought to involve the empty states at the $B$ sites \cite{mayer1981mechanism,zhu2015ferromagnetic,katukuri2020exchange}. 

\begin{table}[!t]
	\centering
	\caption{Table of peak positions and widths in eV extracted from the Rh $L$ edge XAS by fitting a Lorentzian + arctangent lineshape.}
	\begin{tabular}{c c c c c}
	\hline\hline
	{}                       & $E_{L3}$  & $E_{L2}$  & $\Gamma_{L3}$ & $\Gamma_{L2}$ \\ \hline
	Rh acetate               & 3005.5(1) & 3147.7(1) & 2.0(1)        & 1.8(2) \\
    RhCl$_3\cdot x$H$_2$O    & 3005.7(1) & 3148.0(1) & 1.8(1)        & 1.6(2) \\
    NiRh$_2$O$_4$            & 3006.1(1) & 3148.3(1) & 2.0(1)        & 1.7(2) \\
    Rh$_2$O$_3\cdot 5$H$_2$O & 3006.3(1) & 3148.5(1) & 1.9(1)        & 1.7(2) \\
    Ag$_3$LiRh$_2$O$_6$      & 3007.0(1) & 3149.4(1) & 3.0(2)        & 2.4(3) \\
	\hline\hline
	\end{tabular}
	\label{tab:xas}
\end{table}
Fig.~\ref{fig:xas}(b) and Fig.~\ref{fig:xas}(c) show the XAS at the Rh $L_3$ and $L_2$ edges respectively, for \nro{} and reference samples with known valence: Rh acetate (2+), RhCl$_3\cdot x$H$_2$O (3+), Rh$_2$O$_3\cdot 5$H$_2$O (3+), Ag$_3$LiRh$_2$O$_6$ (4+) \cite{bahrami2022first}. The sharp white line peaks correspond to transitions from $2p$ core levels to empty $4d$ states. The Rh$^{2+}$ and Rh$^{3+}$ spectra contain a single peak at each edge, while the Rh$^{4+}$ spectrum contains a shoulder on the low energy side, corresponding to the empty $t_{2g}$ state. As expected, this shoulder is suppressed at the $L_2$ edge \cite{burnus2008xas}. To quantify the Rh valence, we obtain the white line peak position for each compound from a fit to a Lorentzian plus arctangent lineshape \cite{clancy2014dilute,chikara2017charge}. The results of this fit are summarized in table \ref{tab:xas}. The reference compounds for Rh$^{3+}$ show a 0.6~eV (0.5~eV) difference in peak position at the $L_3$ ($L_2$) edge. The higher peak position of Rh$_2$O$_3$ compared to RhCl$_3$ can be attributed to the larger covalency of the Rh-O bond compared to the Rh-Cl bond, giving a more delocalized charge density around the Rh site in Rh$_2$O$_3$ \cite{wu1995xanesRh}. We find that the peak position for \nro{} lies between these two compounds with no significant differences between lineshapes and peak widths. This confirms that despite the strong hybridization, Rh maintains the charge distribution of the 3+ oxidation state with no signs of charge disproportionation, which often occurs in mixed 3d/4d compounds \cite{meyers2014competition,kobayashi2022intervalence,takubo2005xray}.

\begin{figure}[!t]
   \begin{center}
   \includegraphics{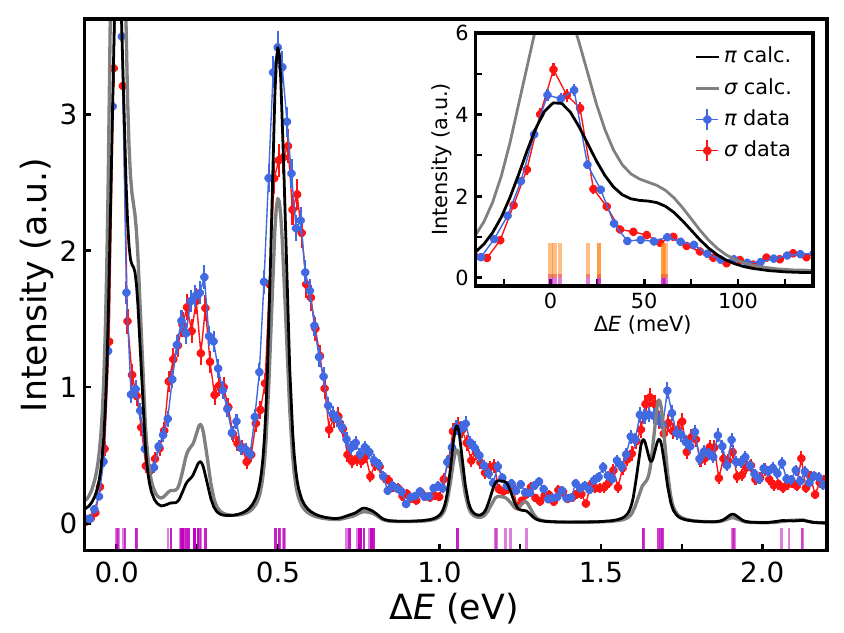}
   \caption{Calculated RIXS spectrum from the two-site model with $t=30$~meV, $\Delta=-600$~meV compared with measured data. Magenta ticks indicate predicted excitation energies.}
   \label{fig:rixs_model_2site}
   \end{center}
\end{figure}
Having confirmed the strong Ni-Rh hybridization and Rh oxidation state in \nro{}, we assign the RIXS peak at 250~meV to a two-site orbital excitation from Rh $t_{2g}$ to Ni $t_2$, corresponding to the transition $\text{Ni }3d^8 + \text{Rh }4d^6 \rightarrow \text{Ni }3d^9 + \text{Rh }4d^5$. This assignment is corroborated by our effective two site model discussed below. Although such metal-metal charge transfer (MMCT) is well known from optical studies of spinel ferrites \cite{fontijn1999consistent,kim2002comparative}, there are few reports of these features seen in RIXS \cite{agui2009intermetallic,chiuzbaian2017spectroscopic,liu2018co2v}. 

A full microscopic description of MMCT in \nro{} is considerably complicated by the 12 fold nearest neighbor Ni-Rh coordination with intermediate Ni-O-Rh bond angles that prohibit any strict orthogonality constraints on hopping pathways. In order to capture the essential features of Ni-Rh hybridization present in the RIXS spectra, we construct a minimal model by adding a single set of filled Rh $t_{2g}$ states to the single-ion model from the previous section (see Supplemental Material \cite{Note1} for details). In this two-site model, the Ni and Rh sites are each subject to Coulomb, crystal field, and spin-orbit interactions, with on-site energy difference $\Delta$ and hopping $t$. Since multiple sites are involved, we now include the monopole part of the Coulomb interaction at each site $U_{\text{Ni}} = 6$~eV and $U_{\text{Rh}} = 3$~eV. We assume an equal hopping from each Ni $t_2$ to each Rh $t_{2g}$ and neglect any hopping from the Ni $e$ levels. The effective crystal field parameters obtained from a single-ion model must also be adjusted in the presence of hopping. As shown in Fig.~\ref{fig:rixs_model_2site}, we find good agreement between data and model for the parameters $\Delta = -600$~meV, $t=30$~meV, $10Dq = -550$~meV, $\lambda = 13$~meV, and $\Delta t_2 = 40$~meV, with the same constraint $\Delta e = 0.8\Delta t_2$. We also set the Rh tetragonal splitting $\Delta t_{2g} = 0$~meV, but the results are mostly unchanged for nonzero $\Delta t_{2g}$. This corresponds to scenario 1 discussed above, but similar results are obtained for scenario 2 (see Supplemental Material \cite{Note1}). We emphasize that the two-site model does not preserve the symmetry of the Ni site, making it ineffective for comparing the two scenarios. This is evident from the inset of Fig. \ref{fig:rixs_model_2site}, where all low-energy levels have non-negligible dipole character.

Despite the simplicity, the minimal model provides a robust qualitative description of the data, reproducing the energy of all observed RIXS excitations over a 2~eV range of energy transfers with effective parameters describing the approximate energy scales of the Ni-Rh hybridization. A more detailed approach should incorporate the empty Rh $e_g$ levels, which may be essential to the long range exchange interaction, and might consider a double cluster model \cite{green2016bond,guedes2019role}, or symmetry-adapted Rh orbitals \cite{haverkort2012multiplet}. In addition, through comparison of high energy inelastic neutron scattering data that is sensitive to optical phonons, with the RIXS spectra, we find that lattice vibrations enter as an essential energy scale coupled to the electronic states in \nro{}. A consideration of electron phonon coupling and vibronic excitations is thus necessary to capture the broad asymmetric lineshape of the RIXS spectra, as discussed in the following section.

\subsection{Electron-phonon coupling} \label{epc}
In orbitally degenerate systems, there is a tendency for strong electron-lattice coupling. In many cases, the degeneracy may be lifted by a static lattice distortion via the Jahn-Teller (JT) mechanism \cite{kocsis2013magnetoelasticity,kocsis2018strong}. Another possibility is the formation of vibronic modes via the dynamical JT effect, where orbital degeneracy is broken by coupling to lattice vibrations \cite{krimmel2005vibronic}. When the JT distortion energy is comparable to the spin-orbit coupling, the system may host a set of spin-orbital-lattice entangled states \cite{streltsov2020JTandSOC}. In \nro{}, the tetragonal distortion does not fully lift the orbital degeneracy \cite{chamorro2018frustrated}. The weak tetrahedral crystal field splitting (in comparison to octahedral) enables the JT energy to be comparable to SOC. Based on these considerations, we expect lattice dynamics to play a key role in the low-lying spin-orbital excitations.



\begin{figure}[!t]
   \begin{center}
   \includegraphics{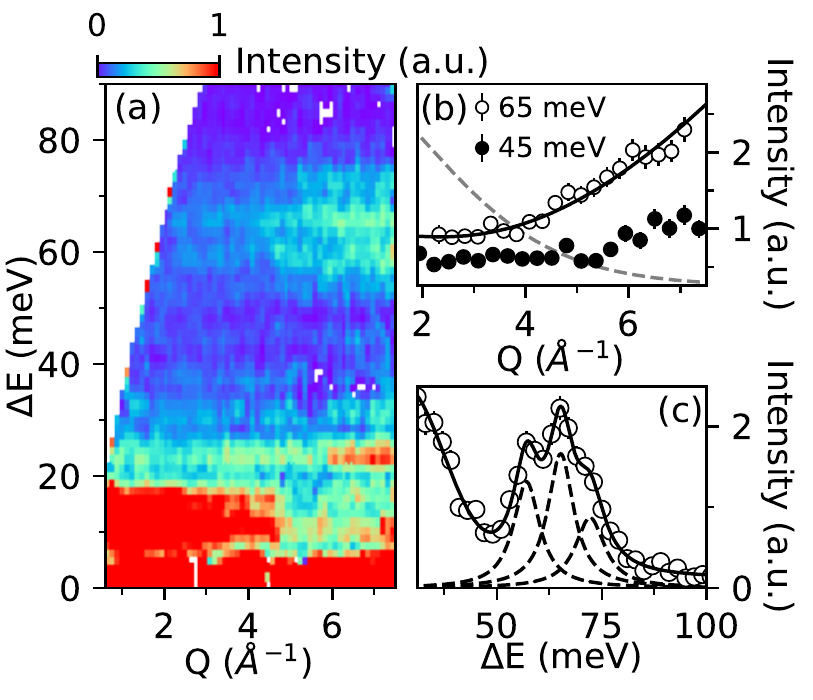}
   \caption{(a) Inelastic neutron scattering spectra $I(Q,E)$ of \nro{} at $T=3.6$~K. Collective spin-orbit excitations are visible around 10~meV, and the highest energy optical phonons appear between 60 and 70 meV. (b) Constant energy cuts through $I(Q,E)$ integrated over $E\pm5$~meV, dashed line shows the momentum dependence of the Ni$^{2+}$ form-factor $|f(Q)|^2$, solid line is a fit to $C|f(Q)|^2 + AQ^2 + B$ as described in the text. (c) Constant momentum-transfer cut integrated over $Q=6\pm1$~\AA$^{-1}$, solid line is a fit to three Lorentzians as described in the text.}
   \label{fig:neutron}
   \end{center}
\end{figure}

Here we consider the effects of electron-phonon coupling on the RIXS spectrum of \nro{}. Although the RIXS spectra may contain contributions from optical phonons, their precise energies are obscured by the relatively coarse energy resolution on the scale of the phonon energies, and the coincidence of optical phonons with low energy electronic excitations. In order to more precisely quantify optical phonon energies in \nro{}, we have re-examined the inelastic neutron scattering data for energies up to 100~meV, covering the full phonon bandwidth. The high energy inelastic neutron scattering data is shown in Fig.~\ref{fig:neutron}(a). 
Below 20~meV, the previously reported collective spin-orbit excitation is visible. As discussed above, this feature was not directly resolved in the RIXS spectra, but accounts for the broadening of the elastic line (inset of Fig.~\ref{fig:rixs}(b) and is captured by the single-ion model. At higher energies, we observe optical phonons, centered around 65~meV, coincident with the intra-$t_2$ excitation in the RIXS spectrum. The quadratically increasing intensity with increasing momentum transfer as shown in Fig.~\ref{fig:neutron}(b) indicates that this signal originates primarly from scattering by phonons, but we also find a component attributable to magnetic scattering. By fitting this cut to $I(Q) = C|f(Q)|^2 + AQ^2 + B$, where $f(Q)$ is the Ni$^{2+}$ magnetic form-factor, $A, B$, and $C$ are constants,  we obtain a good description of the data with the parameters $A=0.04$, $B=0.2$, $C=0.78$. Optical phonon energies were extracted directly from the constant momentum transfer cut in Fig.~\ref{fig:neutron}(c). We found that including three Lorentzian functions centered at 57.2(6), 65.3(5), and 72(1)~meV and with energy linewidths of $\Gamma=$~7.6(1.7), 7.5(2.5), and 8.5(2.0) meV respectively was necessary to adequately describe the data. The phonon linewidths are significantly broadened over the instrumental resolution of $\sim4.3$-meV Gaussian FWHM at 65 meV. These high energy phonons originate from vibrations of Ni coordinating O tetrahedra, and we expect six distinct modes in this energy range for a cubic cell that are further split in the tetragonal phase \cite{ptak2013temperature,kocsis2013magnetoelasticity,wang2003insitu}. Although the broad phonon lineshapes we observed may be accounted for by unresolved phonon mode splittings, it may also indicate phonon damping caused by coupling to other electronic excitations. Indeed, the coinciding energy of these phonons and the intra-$t_2$ crystal field excitation measured by RIXS suggests the possibility of a hybridized vibrational-electronic or ``vibronic'' excitation in \nro{}.

Such electron phonon coupling occurs because the charge distribution of the excited state on Ni repels the surrounding oxygen ions. For the intra $t_2$ excitation at 65~meV, the extra electron in the $d_{xy}$ orbital is partially screened by the resulting $d_{xz}/d_{yz}$ hole. However, the $e\rightarrow t_2$ excitation at 500~meV similarly leaves an extra charge in the $t_2$ levels, with a hole in the $e$ levels. In this case, we expect the screening to be less effective, and thus the $e\rightarrow t_2$ excitation should couple even more strongly to phonons. We can use these considerations to model the lineshape of the 500~meV RIXS peak as a vibronic excitation. Although the single-ion model indicates at least two states comprise this peak and the neutron data shows at least three phonon modes may be involved in the coupling, we will consider a tractable model that includes only a single electronic excitation coupled to a single phonon as such a model is sufficient to capture the essential features of our data. Within this simplified model, we treat the main peak at energy $E_{0}\!=\!496$~meV as a bare $d$-$d$ excitation, or zero-phonon line. This bare dd-excitation is dressed by additional phonon sidebands corresponding to a $d\text{-}d$ excitation plus $n$ phonons of energy $E_{\text{ph}}$. We assume a Lorentzian lineshape of FWHM width $\Gamma$ for each peak separated by energy $E_{\text{ph}}$ with relative intensities given by a Poisson distribution \cite{geondzhian2020large}
\begin{equation} \label{eq:Poisson}
    I_n = e^{-g}\frac{g^n}{n!},
\end{equation}
where $g$ is the dimensionless electron-phonon coupling, which can be interpreted as the mean number of phonons emitted by the excitation. 
\begin{figure}[!t]
   \begin{center}
   \includegraphics{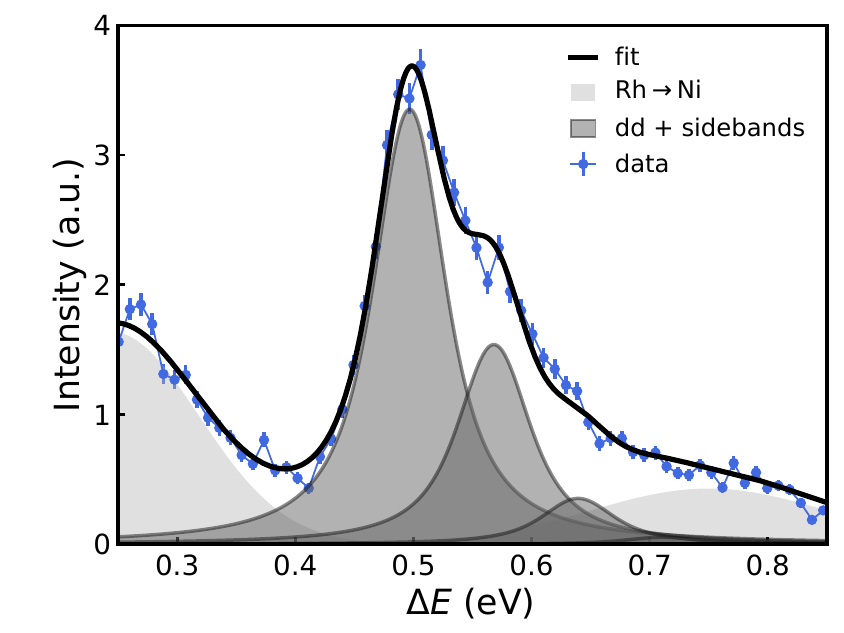}
   \caption{Poisson model for phonon sidebands fit to the 500~meV RIXS peak, including Gaussian fits to the overlapping neighboring peaks. The phonon sideband cross-section was modeled using equation~\ref{eq:Poisson} and Huang-Rhys parameter $g = 0.46(3)$, $\omega_0 = 71(2)$~meV, $E_0 = 497(2)$~meV, and $\Gamma = 60(2)$~meV.}
   \label{fig:rixs_epc}
   \end{center}
\end{figure}
Fig.~\ref{fig:rixs_epc} shows a fit of this minimal model to the 500~meV RIXS peak for $E_0=496(1)$~meV, $E_{\text{ph}}=68(2)$~meV, $\Gamma=70(2)$~meV, and $g=0.45(3)$ providing an excellent description of the data. The large value of $\Gamma$ compared to the phonon energies suggests that there are many overlapping excitations within each phonon sideband. This is consistent with the single-ion model for scenario 1 that gives two crystal field excitations, at 492 and 517~meV. Both of these may couple to optical phonons with energies between 57 and 72~meV. A more complete model for electron phonon coupling in \nro{} would consider the separate $d$-$d$ excitations and their coupling to multiple phonons, but the resolution of our measurement is not sufficient to constrain such a model. 

\section{Discussion and Conclusion} \label{conclusion}
We have characterized the site-specific local electronic structure of \nro{} using resonant inelastic x-ray scattering and x-ray absorption spectroscopy. We have compared two possible scenarios for the tetragonal splitting within a single ion model and showed that $\Delta t_2 > 0$ is more likely than $\Delta t_2 < 0$, and estimated that $\Delta t_2 = 70$~meV and $\lambda = 13$~meV. These parameters are the most relevant to modeling the magnetism in \nro{} \cite{li2019spinorbital,das2019nirh2o4}. The crystal field splittings are in agreement with DFT calculations that determined $\Delta t_2 = 100$~meV and $\lambda = 10$~meV from NMTO downfolding. The single ion model also required a 50\% reduction of the $F_{dd}$ Slater parameters, suggesting a significant degree of covalency  in \nro{}.

The O $K$ edge XAS data suggests a significant degree of hybridization between O $p$ states and empty metal $d$ states, which can only be explained by a metal-metal charge transfer between adjacent Ni and Rh sites. The Rh $L$ edge XAS confirmed that the Rh ions maintain the nominal $3+$ oxidation state, despite the strong hybridization and tendency toward charge disproportionation in related systems. By extending the RIXS model to include metal-metal charge transfer between Ni and Rh sites as parameterized by an effective hopping $t\!=\!30$~meV, we captured Rh-Ni two-site excitations observed at 250 and 750~meV in the RIXS spectrum. Such an explicit demonstration of the failure of a single ion model and requirement for metal-metal charge transfer in \nro{} highlights the importance of metal-metal hybridization in mixed 3d-4d/5d compounds in general. This hybridization can affect both the magnetic degrees of freedom and exchange interactions so should be an essential consideration in the design of novel magnetic states in materials such as spinels, double perovskites \cite{sahadasgupta2020double,lee2018hybridized}, or $A_2$Mo$_3$O$_8$ compounds \cite{morey2019ni2mo3o8,park2021bandmott}.

A detailed analysis of the RIXS lineshape also revealed that lattice vibrations influence the magnetism in \nro{} through strong electron-phonon coupling. Inelastic neutron scattering reveals multiple optical phonons overlapping in energy with the intra-$t_2$ excitation, suggesting a hybridized orbital-lattice excitation between 60 and 70~meV. This effect was observed in the RIXS spectra as a phonon-dressed crystal field excitation at higher energy. Our results provide quantitative constraints on the key parameters for modeling the single-ion ground state and low-lying excitations, as well the long-range superexchange mechanism in \nro{}. We also demonstrate the importance of additional degrees of freedom, namely covalency and phonons, which can alter magnetic ground states \cite{xu2016covalency}. Our results also demonstrate the use of RIXS for probing spin-orbit entangled states in $3d$ transition metal compounds \cite{huang2022resonant} and for probing hybridized states in mixed 3d-4d/5d compounds \cite{jin2022magnetic}.

In addition to the predicted phenomena associated with spin-1 frustrated diamond lattice antiferromagnets, we propose that \nro may host a variety of novel magneto-elastic and magneto-optical effects due to its rich spectrum of low-energy spin-orbital-lattice excitations \cite{kocsis2013magnetoelasticity,kocsis2018strong,smerald2019giant,amelin2020thz}. Future studies on \nro{} could use pressure, magnetic fields or chemical substitution to explore the predicted phase diagram and observe quantum critical phenomena \cite{chen2017quantum,buessen2018quantum,li2019spinorbital}. These studies would greatly benefit from the synthesis of single crystals or thin films. Should such crystals become available, another potentially fruitful route would be ultrafast optics to study the spin-orbital and crystal field excitations with the lattice out-of-equilibrium \cite{marciniak2021vibrational,afanasiev2021ultrafast,ergeccen2022magnetically}. By resonantly exciting the lattice, it may be possible to stabilize an excitonic condensate of $J=1$ moments \cite{li2019spinorbital}.

\section*{Acknowledgements} \label{acknowledgements}

We thank Arun Paramekanti, Pat Clancy, and Hlynur Gretarsson for useful discussions. We are grateful to Matthew Stone for assistance in collecting data at ORNL. We thank Fazel Tafti and Faranak Bahrami for providing the Ag$_3$LiRh$_2$O$_6$ reference sample for XAS. Work at Brown University was supported by the U.S. Department of Energy, Office of Science, Office of Basic Energy Sciences, under Award Number DE-SC002165. TMM and JC acknowledge support from the Institute for Quantum Matter, an Energy Frontier Research Center funded by the U.S. Department of Energy, Office of Science, Office of Basic Energy Sciences, under Award DE-SC0019331.The work of L.G. and M.M. at Georgia Tech (inelastic neutron scattering measurements) was supported by NSF-DMR-1750186. This research used beamline 2-ID of the National Synchrotron Light Source II, a U.S. Department of Energy (DOE) Office of Science User Facility operated for the DOE Office of Science by Brookhaven National Laboratory under Contract No. DE-SC0012704. Use of the Advanced Photon Source at Argonne National Laboratory was supported by the U. S. Department of Energy, Office of Science, Office of Basic Energy Sciences, under Contract No. DE-AC02-06CH11357.A portion of this research used resources at the Spallation Neutron Source, a DOE Office of Science User Facility operated by the Oak Ridge National Laboratory.


%

\end{document}


\title{\texorpdfstring{Supplemental Material \\ Electronic structure of the frustrated diamond lattice magnet \nro}{}}

\author{B. Zager}
\affiliation{Department of Physics, Brown University, Providence, Rhode Island 02912, United States}
\author{J. R. Chamorro}
\affiliation{Department of Chemistry, Johns Hopkins University, Baltimore, MD, USA}
\affiliation{Institute for Quantum Matter, William H. Miller III Department of Physics and Astronomy, Johns Hopkins University, Baltimore, MD, USA}
\author{L. Ge}
\affiliation{School of Physics, Georgia Institute of Technology, Atlanta, Georgia 30332, USA}
\author{V. Bisogni}
\affiliation{National Synchrotron Light Source II, Brookhaven National Laboratory, Upton, New York 11973, USA}
\author{J. Pelliciari}
\affiliation{National Synchrotron Light Source II, Brookhaven National Laboratory, Upton, New York 11973, USA}
\author{J. Li}
\affiliation{National Synchrotron Light Source II, Brookhaven National Laboratory, Upton, New York 11973, USA}
\author{G. Fabbris}
\affiliation{Advanced Photon Source, Argonne National Laboratory, Lemont, Illinois 60439, USA}
\author{T. M. McQueen}
\affiliation{Department of Chemistry, Johns Hopkins University, Baltimore, MD, USA}
\affiliation{Institute for Quantum Matter, William H. Miller III Department of Physics and Astronomy, Johns Hopkins University, Baltimore, MD, USA}
\affiliation{Department of Materials Science and Engineering, Johns Hopkins University, Baltimore, MD, USA}
\author{M. Mourigal}
\affiliation{School of Physics, Georgia Institute of Technology, Atlanta, Georgia 30332, USA}
\author{K. W. Plumb}
\affiliation{Department of Physics, Brown University, Providence, Rhode Island 02912, United States}

\date{\today}

\maketitle
In the following supplemental material we provide additional details on modeling the RIXS cross-section within a single-site model that accounts for tetrahedrally coordinated Ni$^{2+}$, and a two site model that accounts from Ni-Rh hybridization. We also provide additional figures that show a detailed exploration of the parameter space for the single and two site RIXS models compared against our data.

\section{RIXS simulations} \label{ED}





We compute the RIXS cross section \cite{ament2011rixs} in the dipole approximation from the eigenstates of model Hamiltonians obtained from exact diagonalization using the EDRIXS software package \cite{wang2019edrixs}. The calculated spectra are convolved with a 31 meV FWHM Gaussian to account for the experimental energy-loss resolution. The spectra are calculated for $2\theta = 90\degree$ scattering angle corresponding to $Q = 0.61\;\text{\AA}^{-1}$, powder averaged over 100 random orientations of $\bm{Q}$ uniformly sampled over the surface of a sphere.

\subsection{Single-site model}

In the single-ion model, we consider a Ni$^{2+}$ ion with Coulomb interaction, spin-orbit coupling (SOC), and a tetrahedral crystal field with tetragonal distortion. In the initial/final state, this corresponds to a $3d^8$ configuration with 45 states. In the intermediate state, this corresponds to a $2p^53d^9$ configuration with 60 states. The Hamiltonian for the initial state ($3d^8$) is given by
\begin{equation}
    \HH_i^{} = \HH_U^{} + \HH_{\text{CF}}^{} + \HH_{\text{SOC}}^{},
\end{equation}
where the Coulomb interaction is given by
\begin{equation}
    \HH_U^{} = \sum\limits_{\alpha\beta\gamma\delta\sigma\sigma'} 
    U_{\alpha\sigma,\beta\sigma',\gamma\sigma',\delta\sigma}
    f_{\alpha\sigma}^{\dagger}f_{\beta\sigma'}^{\dagger}
    f_{\gamma\sigma'}^{}f_{\delta\sigma}^{}
\end{equation}
where $f^{\dagger}$ and $f$ are creation and annihilation operators and $U$ is the rank-4 Coulomb tensor. This describes the Coulomb interaction between orbitals indexed by $\alpha$, $\beta$, $\gamma$, and $\delta$, and spin indexed by $\sigma$ and $\sigma'$. For an explicit representation of $U$, see \cite{cowan1981theory,haverkort2005spin}. Here the relevant parameters in $\HH_U$ are the Slater integrals $F^k$ and $G^k$, which are calculated from Cowan's code \cite{cowan1981theory} using the Hartree-Fock method. To agree with the data, the Slater integrals are scaled down to account for the reduced intra-atomic repulsion due to covalency in the solid. For the Coulomb interaction between two $d$ orbitals in the single-ion model, the only relevant Slater integrals are $F_{dd}^2$ and $F_{dd}^4$. The spin-orbit coupling is given by
\begin{equation}
    \HH_{\text{SOC}}^{} = \lambda \bm{L}\cdot\bm{S}.
\end{equation}
The crystal field $\HH_{\text{CF}}$ is given by a $5\times 5$ matrix that is diagonal in the $(d_{z^2},d_{xz},d_{yz},d_{x^2-y^2},d_{xy})$ basis, with elements
\begin{align}
    E_{z^2} &= 6Dq + \tfrac{2}{5}\Delta t_2 - \tfrac{4}{5}\Delta e \\
    E_{xz} &= -4Dq - \tfrac{3}{5}\Delta t_2 + \tfrac{1}{5}\Delta e \\
    E_{yz} &= -4Dq - \tfrac{3}{5}\Delta t_2 + \tfrac{1}{5}\Delta e \\ 
    E_{x^2} &= 6Dq + \tfrac{2}{5}\Delta t_2 + \tfrac{1}{5}\Delta e \\
    E_{xy} &= -4Dq + \tfrac{2}{5}\Delta t_2 + \tfrac{1}{5}\Delta e.
\end{align}
To include the spin degree of freedom, we use a $10\times 10$ matrix with equal crystal field splitting for each spin the same orbital. The intermediate state ($2p^53d^9$) Hamiltonian is given by
\begin{equation}
    \HH_n^{} = \HH_i^{} + \HH_p^{} + \HH_{pd}^{}
\end{equation}
where $\HH_i$ has the same form as above but for a $3d^9$ configuration. $\HH_p^{}$ is the Ni $2p$ Hamiltonian which includes an on-site energy difference and spin-orbit coupling with $\lambda_{2p} = 11.507$~eV. $\HH_{pd}^{}$ is the Coulomb interaction between the core hole and valence electrons, which is parameterized by the Slater integrals $F_{pd}^2$, $G_{pd}^1$, and $G_{pd}^3$. 


\subsection{Two-site model}


To consider the effects of Ni-Rh hybridization, we construct a two-site model including the Ni$^{2+}$ site described above, as well as a Rh site with filled $t_{2g}$ orbitals. We neglect the empty Rh $e_g$ orbitals to reduce the size of the Hilbert space. The initial/final state in this model is a mixture of the configurations $3d_{}^8t_{2g}^6$, $3d_{}^9t_{2g}^5$, and $3d_{}^{10}t_{2g}^4$, with 120 total states. The intermediate state is a mixture of the configurations $2p^5 3d_{}^9t_{2g}^6$ and $2p^5 3d_{}^{10}t_{2g}^5$, with 96 total states. We ignore any interactions between the core hole and the Rh site. The initial/final state Hamiltonian is given by
\begin{equation}
    \HH_i^{} = \HH_{\text{Ni}} + \HH_{\text{Rh}} + V
\end{equation}
where $\HH_{\text{Ni}}$ ($10\times 10$) is equivalent to $\HH_i$ from the single-ion model,
$\HH_{\text{Rh}}$ ($6\times 6$) is the Hamiltonian for the Rh $t_{2g}$ levels, and $V$ is the hybridization term which mixes the single-ion states, given by
\begin{equation}
    V = 
    \begin{pmatrix}
        0 & T^{\dagger} \\
        T^{} & 0
    \end{pmatrix},
    \qquad
    T = 
    \begin{pmatrix}
        0 & t & t & 0 & t \\
        0 & t & t & 0 & t \\
        0 & t & t & 0 & t
    \end{pmatrix}.
\end{equation}
This equally mixes each of the 5 Ni $d$ orbitals with each of the 3 Rh $t_{2g}$ orbitals. Since this model considers multiple sites, we now include the monopole term in the Coulomb interaction on each site, $F^0 = U + \frac{2}{63}(F^2 + F^4)$, where $U_{\text{Ni}} = 6$~eV, $U_{\text{Rh}} = 3$~eV. The intermediate state Hamiltonian $\HH_n$ is equivalent to that of the single-ion model, since we neglect any explicit interaction between the core-hole and the Rh site. We also include on-site energies $E_{\text{Ni}}$ and $E_{\text{Rh}}$. The effective charge transfer energy $\Delta$ gives the energy of the $\text{Ni}\;3d^9\;\text{Rh}\;4d^5$ configuration relative to $\text{Ni}\;3d^8\;\text{Rh}\;4d^6$ before any splitting. We use this to determine the on-site energies $E_{\text{Ni}}$ and $E_{\text{Rh}}$ for the initial state by solving the linear system of equations
\begin{align}
    \nNi\ENi + \nRh\ERh + \tfrac{1}{2}\nNi(\nNi-1)\UNi + \tfrac{1}{2}\nRh(\nRh{}-1)\URh &= 0 \\
    (\nNi+1)\ENi + (\nRh-1)\ERh + \tfrac{1}{2}\nNi(\nNi+1)\UNi + \tfrac{1}{2}(\nRh-1)(\nRh-2)\URh &= \Delta.
\end{align}
In the intermediate state, we obtain the on-site energies of the Ni and Rh $d$ orbitals $E_{\text{Ni}}'$ and $E_{\text{Rh}}'$, as well as the Ni $2p$ orbitals $E_p$, from the system of equations
\begin{align}
    6E_p + \nNi E'_{\text{Ni}}    + \nRh E'_{\text{Rh}}    + \tfrac{1}{2}\nNi(\nNi-1)\UNi + \tfrac{1}{2}\nRh(\nRh-1)\URh + 6\nNi U_{pd} &= 0 \\
    6E_p + (\nNi+1)E'_{\text{Ni}} + (\nRh-1)E'_{\text{Rh}} + \tfrac{1}{2}(\nNi+1)\nNi\UNi + \tfrac{1}{2}(\nRh-1)(\nRh-2)\URh + 6(\nNi+1)U_{pd} &= \Delta \\
    5E_p + (\nNi+1)E'_{\text{Ni}} + \nRh E'_{\text{Rh}}    + \tfrac{1}{2}(\nNi+1)\nNi\UNi + \tfrac{1}{2}\nRh(\nRh-1)\URh + 5(\nNi+1)U_{pd} &= 0.
\end{align}

\clearpage

\section{Additional figures for RIXS models} \label{}

\begin{figure}[!h]
   \begin{center}
   \includegraphics{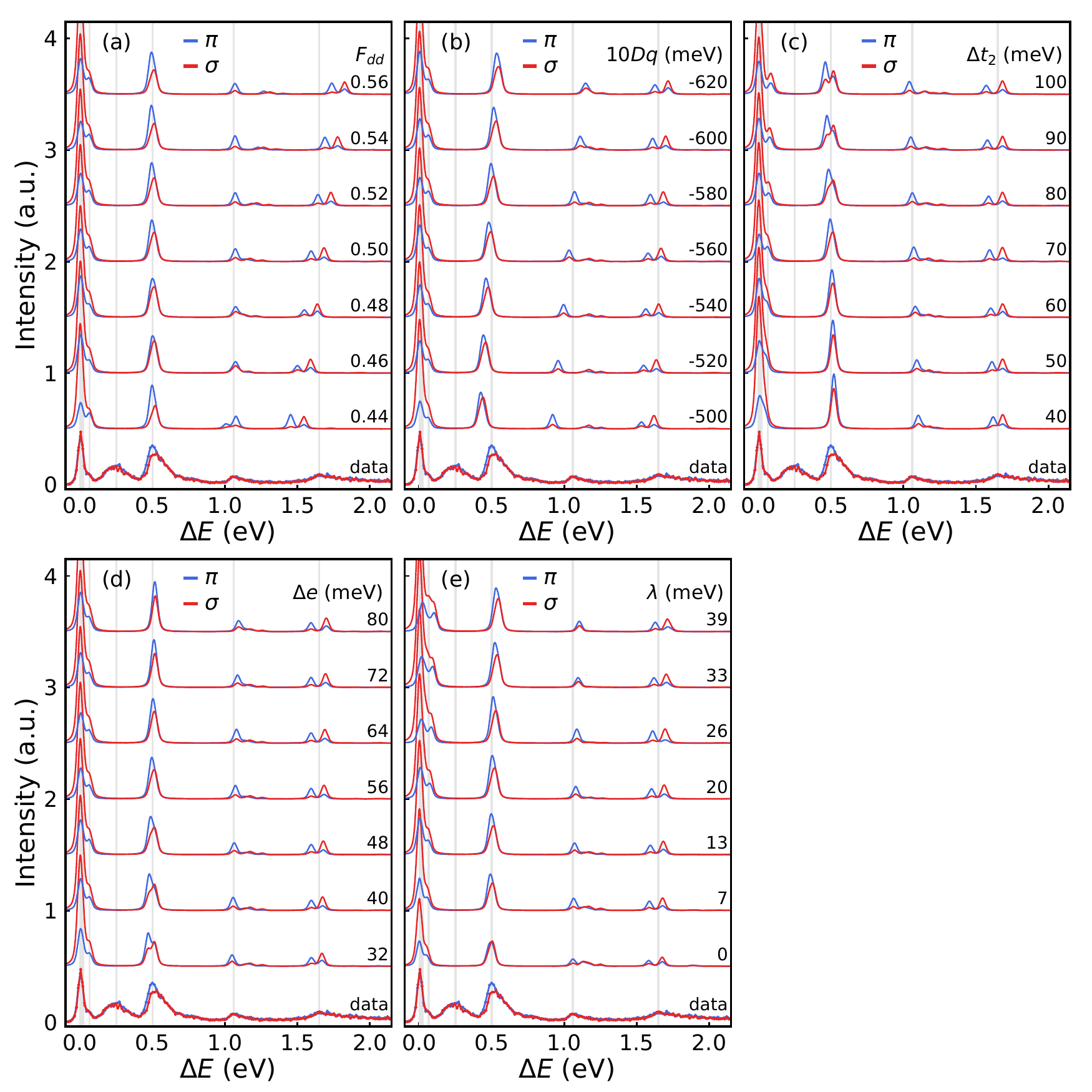}
   \caption{Calculated RIXS spectra for scenario 1, varying each parameter with all other parameters fixed to the optimal values. Vertical gray lines mark the energies of the measured peaks.
   (a) $F_{dd}$
   (b) $10Dq$
   (c) $\Delta t_2$
   (d) $\Delta e$
   (e) $\lambda$
   }
   \label{fig:rixs_sweep_1}
   \end{center}
\end{figure}

\begin{figure}[!h]
   \begin{center}
   \includegraphics[scale=0.9]{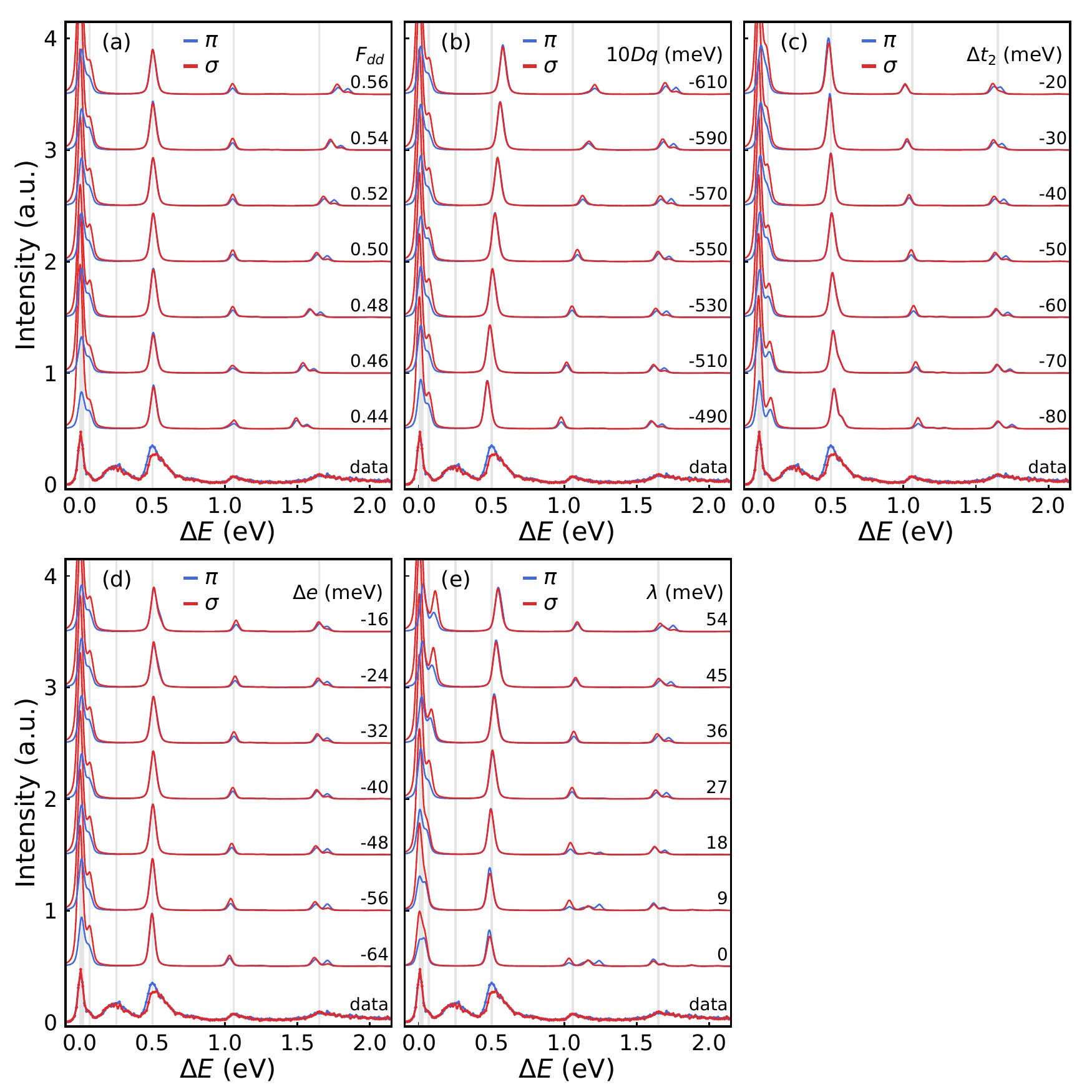}
   \caption{Calculated RIXS spectra for scenario 2, varying each parameter with all other parameters fixed to the optimal values. Vertical gray lines mark the energies of the measured peaks.
   (a) $F_{dd}$
   (b) $10Dq$
   (c) $\Delta t_2$
   (d) $\Delta e$
   (e) $\lambda$
   }
   \label{fig:rixs_sweep_2}
   \end{center}
\end{figure}

\begin{figure}[!h]
   \begin{center}
   \includegraphics[scale=0.9]{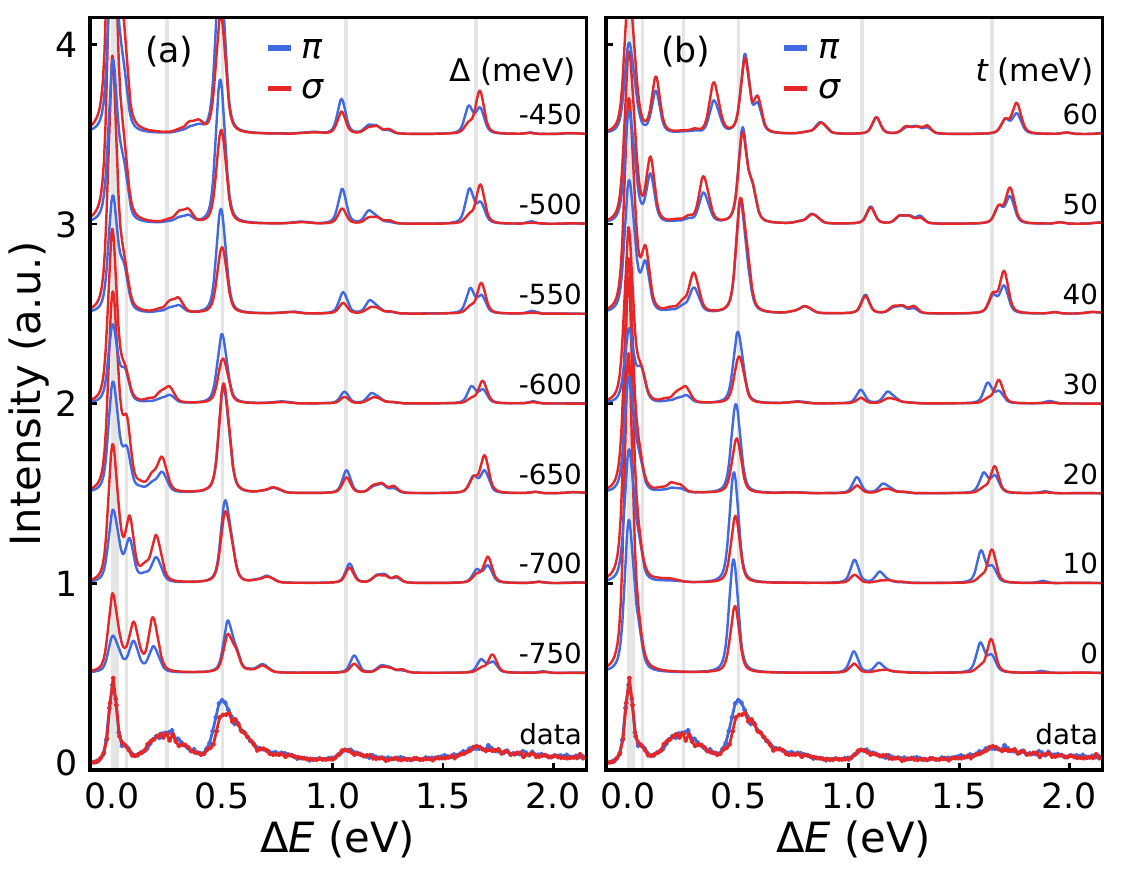}
   \caption{Calculated RIXS spectra for scenario 1 using the two-site model, varying each parameter with all other parameters fixed to optimal values. 
   (a) $\Delta$
   (b) $t$
   }
   \label{fig:rixs_sweep_2site_1}
   \end{center}
\end{figure}

\begin{figure}[!h]
   \begin{center}
   \includegraphics[scale=0.85]{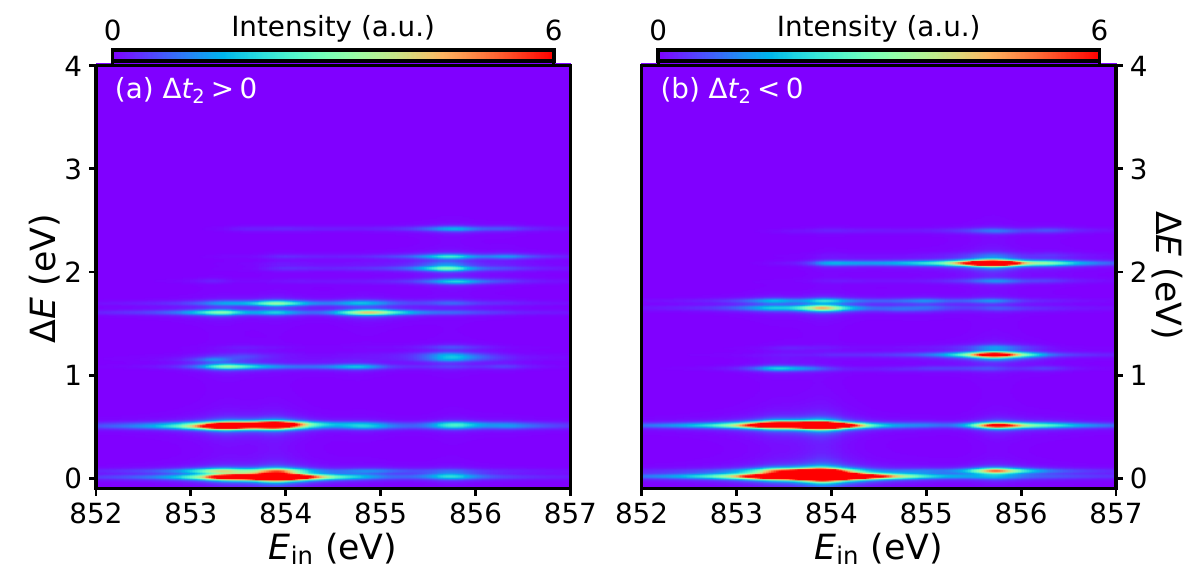}
   \caption{Calculated RIXS maps for $\pi$-polarization using the single-ion model. (a) scenario 1 (b) scenario 2}
   \label{fig:rixs_map_calc}
   \end{center}
\end{figure}

\begin{figure}[!h]
   \begin{center}
   \includegraphics[scale=0.85]{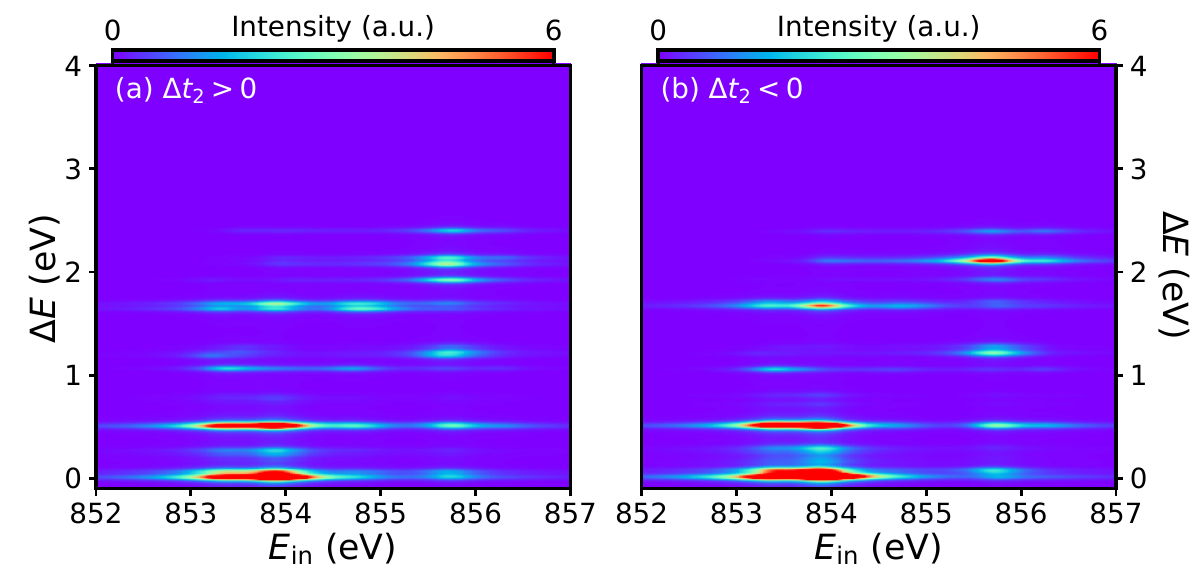}
   \caption{Calculated RIXS maps for $\pi$-polarization using the two-site model. (a) Scenario 1 (b) Scenario 2}
   \label{fig:rixs_map_calc_2site}
   \end{center}
\end{figure}

In this section, we show the dependence of the calculated RIXS spectra on the model parameters, as well as calculated RIXS energy maps showing the incident energy dependence. Fig. \ref{fig:rixs_sweep_1} and Fig. \ref{fig:rixs_sweep_2} show the calculated spectra from the single-ion model for scenarios 1 and 2 respectively, demonstrating the dependence on the parameters $F_{dd}$, $10Dq$, $\Delta t_2$, $\Delta e$, and $\lambda$. In each plot, the given parameter is varied with all other parameters fixed to the optimal values reported in the main text. Fig. \ref{fig:rixs_sweep_2site_1} shows the dependence of the two-site model on the parameters $\Delta$ and $t$. Fig. \ref{fig:rixs_map_calc} and \ref{fig:rixs_map_calc_2site} show the incident energy dependence of the RIXS spectra for the single-ion and two-site models respectively. The energy loss extends up to 4 eV to allow a direct comparison to the measured RIXS map shown in the main text.

\clearpage

%